\documentclass[letter,twocolumn,amsmath,amssymb,superscriptaddress]{jpsj3}
\usepackage{txfonts,amsmath,amssymb}
\usepackage{graphicx}
\usepackage{amssymb}
\usepackage{amsmath}
\usepackage{bm}
\usepackage{color}

\newcommand{\nc}{\newcommand}
\nc{\mb}{\mbox}

\newcommand{\vB}{\bm{B}}
\newcommand{\vm}{\bm{m}}
\newcommand{\vM}{\bm{M}}
\newcommand{\vk}{\bm{k}}
\newcommand{\vs}{\bm{\sigma}}

\newcommand{\vQ}{\bm{Q}}

\newcommand{\va}{\bm{a}}
\newcommand{\vb}{\bm{b}}
\newcommand{\vc}{\bm{c}}
\newcommand{\vn}{\bm{n}}

\nc{\be}{\begin{equation}}
\nc{\ee}{\end{equation}}
\nc{\bea}{\begin{eqnarray}}
\nc{\eea}{\end{eqnarray}}
\nc{\bean}{\begin{eqnarray*}}
\nc{\eean}{\end{eqnarray*}}

\title{Anomalous Hall Effect and Spontaneous Orbital Magnetization in Antiferromagnetic Weyl Metal}

\author{Naohiro \textsc{Ito}$^1$, 
Kentaro \textsc{Nomura}$^1$\thanks{nomura@imr.tohoku.ac.jp}
}
\inst{$^1$Institute for Materials Research, Tohoku University, Sendai 980-8577
}

\kword{%
anomalous Hall effect, orbital magnetization, chiral antiferromagnet, Weyl semimetal
}

\abst{
We study the anomalous Hall effect and the orbital magnetization
 in chiral antiferromagnets,
 constructing a simple tight-binding model on a stacked Kagome lattice structure with spin-orbit coupling and the exchange interaction between the localized spins and itinerant electrons.
It is shown that the chiral antiferromagnet with spin-orbit coupling is an orbital ferromagnet with the easy-plane anisotropy.
The relation between the orbital magnetization and the configuration of the Weyl points are discussed.
}

\begin{document}
\maketitle


{\it Introduction}---
The anomalous Hall effect is observed in ferromagnetic materials with spin-orbit coupling\cite{Nagaosa2010,Xiao2010}.
It is known that the anomalous Hall effect is originated from extrinsic scattering by spin-orbit coupled scatterers or intrinsic contribution from the spin-orbit coupled electronic band structure\cite{Nagaosa2010}.
In most cases the Hall resistivity is expressed in the following form
\begin{equation}
 \rho_{xy}=R_0B_z^{\rm ext}+R_sM_z^{\rm spin},
\label{rho_xy}
\end{equation}
where $B_z^{\rm ext}$ is an external magnetic field and $M_z^{\rm spin}$ is the spin magnetization.
At zero field, the Hall conductivity of a ferromagnetic metal is proportional to the perpendicular component of  the spin magnetization\cite{Nagaosa2010},
\begin{equation}
 \sigma_{ij}^{\rm AHE} \propto \epsilon_{ijk} M_k^{\rm spin}.
\label{sigma_xy_FM}
\end{equation}
 Because of this relation, the anomalous Hall effect is used to detect the magnetization in ferromagnetic materials.
In magnetic materials with noncoplanar spin configurations a finite scalar spin chirality acts as a fictitious magnetic field in real space for the conduction electrons and contributes to the anomalous Hall effect even without spin-orbit coupling\cite{Nagaosa2010}.

The anomalous Hall effect in antiferromagnets has been theoretically studied.
\cite{Kleiner1966,
Maranzana1967,
Shindou2001,
Mandal2012,
Chen2014,
Kubler2014,
Sekine2016,
Yang2017,
Suzuki2017} 
Recently, the anomalous Hall effect was experimentally observed at room temperature in noncollinear antiferromagnets Mn$_3Z$, where $Z=$ Sn, Ge, and Ir
\cite{Nakatsuji2015,
Nayak2015,
Kiyohara2016,Li2017}.
The anomalous Hall conductivity is as large as that in conventional ferromagnetic metals such as Iron and Cobalt even though the net spin magnetization is negligibly small\cite{Nakatsuji2015}.
Mn$_3Z$ has the layered hexagonal lattice structure. Inside a layer, Mn atoms form a Kagome-type lattice with mixed triangles and hexagons and $Z$ (Sn, Ge, and Ir) atoms are located at the center of these hexagons. 
All Mn moments are arranged inside the Kagome plane forming triangular spin structures, referred to  chiral antiferromagnets(CAF). 
Since the scalar spin chirality is zero in antiferromagnets, the mechanism of anomalous Hall effect is essentially different from that in magnetic materials with noncoplanar spin structures.

Electronic structures in chiral antiferromagnets have been studied using first principle calculation
\cite{Chen2014,Kubler2014,Yang2017,Suzuki2017} 
, and the existence of multiple Weyl points has been observed\cite{Yang2017}.
It has been also shown that the mechanism of the anomalous Hall effect in Mn$_3Z$ can be characterized by the cluster extension of octupole moments in the same manner as that in ferromagnets by the magnetization of dipole moment\cite{Suzuki2017}.
Up to now, however, a simple tight-binding model has not been studied in detail.
In this paper we construct an effective tight-binding Hamiltonian with the exchange interaction and spin-orbit coupling on a stacked Kagome lattice.
We show that the system has a spontaneous orbital magnetization even though net spin magnetization is zero. The anomalous Hall conductivities, $\sigma_{xy}$, $\sigma_{yz}$, and $\sigma_{zx}$ are computed using the Kubo formula and connected with the orbital magnetization.
The existence of Weyl points plays an important role in the anomalous Hall effect. We discuss the relation between the orbital magnetization and the configuration of the Weyl points.



\begin{figure}[t]
 \leavevmode\includegraphics[width=1.0\hsize]{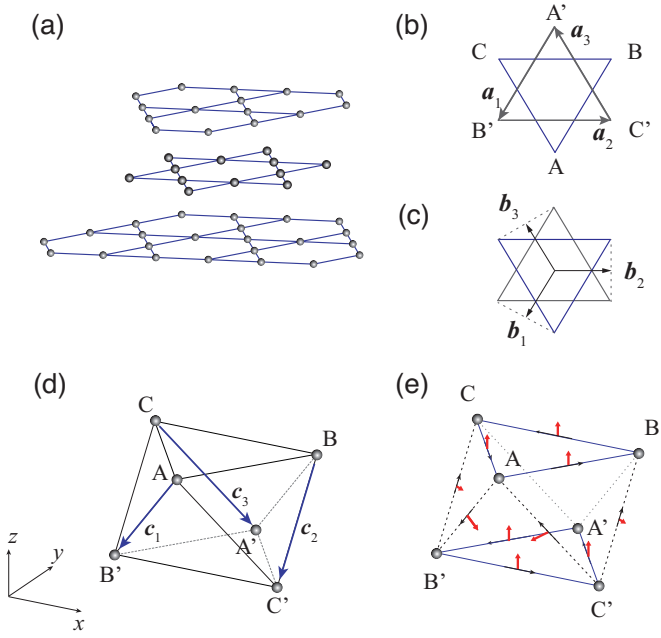} 
 \caption{
 (a) Structure of layered hexagonal lattice structure in Mn$_3Z$ ($Z$=Sn, Ge, and Ir). Each layer forms Kagome lattice of Mn atoms.  $Z$ atoms are located at the center of hexagons.
(b) Nearest-neighbor vectors $\va_i$ in the Kagome plane of Mn atoms.
(c) $\vb_i$ vectors characterize the symmetry of the spin-orbit interaction. 
(d) Nearest-neighbor vectors $\vc_i$ between neighboring Kagome planes. 
(e) Vectors in the spin-orbit coupling term in Eq. (\ref{eq:SOC}).
}
 \label{fig:lattice01}
\end{figure}

{\it Model Hamiltonian}---
In the following we construct a single-orbital tight-binding model on the layered hexagonal lattice (the stacked Kagome lattice structure) of Mn atoms, neglecting electronic degrees of freedom on $Z$ sites, as shown in Fig. \ref{fig:lattice01} (a).
We start from the hopping term
\begin{equation}
H_0=-\sum_{\langle i,j\rangle}t_{ij}c_{i\sigma}^{\dag}c_{j\sigma}^{},
\end{equation}
where $c^{\dag}_{i\sigma}$ creates an electron with spin $\sigma$ ($=\,\uparrow$ or $\downarrow$) at the $i$th site of the stacked Kagome layers, and $\langle i,j\rangle$ denotes nearest neighbors.
As Fig. \ref{fig:lattice01} shows, a unit cell of the crystal consists of six atoms located at 
$A=(0,0,\sqrt{\frac{2}{3}}c)$, $B=(\frac{a}{2},\frac{\sqrt{3}a}{2},\sqrt{\frac{2}{3}}c)$,  $C=(-\frac{a}{2},\frac{\sqrt{3}a}{2},\sqrt{\frac{2}{3}}c)$ in one layer and 
$A'=(0,\frac{2a}{\sqrt{3}},0)$, $B'=(-\frac{a}{2},\frac{a}{2\sqrt{3}},0)$,  $C'=(\frac{a}{2},\frac{a}{2\sqrt{3}},0)$ in another layer. Here $a$ and $c$ are lattice spacing intralayer and interlayer, respectively.
The intralayer nearest-neighbor vectors are given by 
$\va_1=(-\frac{a}{2},-\frac{\sqrt{3}a}{2},0)$, $\va_2=(a,0,0)$, $\va_3=(-\frac{a}{2},\frac{\sqrt{3}a}{2},0)$,
and interlayer nearest-neighbor vectors are
$\vc_1=(-\frac{a}{2},-\frac{a}{2\sqrt{3}},\sqrt{\frac{2}{3}}c)$, $\vc_2=(0,-\frac{a}{\sqrt{3}},-\sqrt{\frac{2}{3}}c)$, $\vc_3=(\frac{a}{2},\frac{a}{2\sqrt{3}},-\sqrt{\frac{2}{3}}c)$ as shown in Fig. \ref{fig:lattice01}.
In the following we set $t_{ij}=t_0$ (between nearest neighbor, otherwise zero) and $a=c=1$ for simplicity.

\begin{figure}[b]
 \leavevmode\includegraphics[width=1.0\hsize]{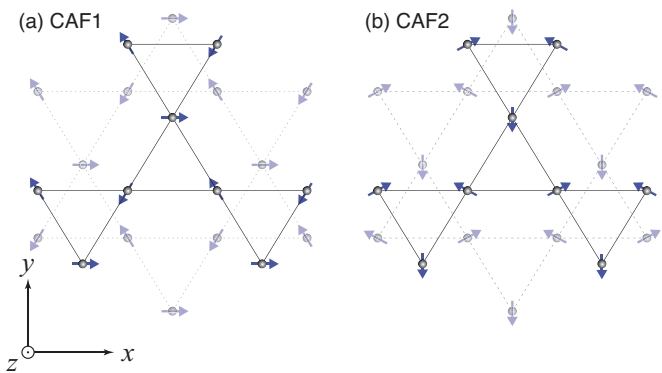} 
 \caption{
Configuration of local spins on Mn atoms (a) CAF1 and (b) CAF2. Spin configurations are experimentally realized depending on the direction of magnetic fields along $x$ and $y$, respectively.
 }
 \label{fig:lattice02}
\end{figure}

Next we introduce spin-dependent nearest-neighbor hopping which is one of ingredients for the anomalous Hall effect. 
In the stacked Kagome lattice inversion symmetry is locally broken. In a process of nearest neighbor hopping an electron feels an electric field perpendicular to the direction of hopping that gives rise to the spin-orbit coupling term of the form
\begin{equation}
H_{\rm so}=it_{\rm so}\sum_{<i,j>}\bm{n}_{ij}\cdot c_{i\sigma}^{\dag}\vs_{\sigma\sigma'} c_{j\sigma'}^{},
\label{eq:SOC}
\end{equation}
where $t_{\rm so}$ is the strength of spin-orbit coupling and $\bm{n}_{ij}\, (=-\bm{n}_{ji})$ is the unit vectors perpendicular to the directions of hopping and the local electric fields.
In a Kagome layer $\vn_{BA}=\vn_{CB}=\vn_{AC}
=\vn_{B'A'}=\vn_{C'B'}=\vn_{A'C'}
=(0,0,1)$, 
while $\vn_{A'B}=\vn_{B'A}=2\vb_1\times \vc_{1}=(\frac{1}{\sqrt{2}},-\frac{1}{\sqrt{6}},-\frac{1}{\sqrt{3}})$ along $\vc_1$, 
$\vn_{B'C}=\vn_{C'B}=2\vb_2\times \vc_{2}=(0,\sqrt{\frac{2}{3}},-\frac{1}{\sqrt{3}})$ along $\vc_2$,
and 
$\vn_{C'A}=\vn_{A'C}=2\vb_3\times \vc_{3}=(-\frac{1}{\sqrt{2}},-\frac{1}{\sqrt{6}},-\frac{1}{\sqrt{3}})$ along $\vc_3$,
where
 $\vb_1=(-\frac{1}{4},-\frac{\sqrt{3}}{4},0)$, $\vb_2=(\frac{1}{2},0,0)$, $\vb_3=(-\frac{1}{4},\frac{\sqrt{3}}{4},0)$ as shown in Fig. \ref{fig:lattice01}(b). All unit vectors, ${\bm n}_{ij}$, are illustrated in Fig. \ref{fig:lattice01}(e). 

\begin{figure}[t]
 \leavevmode\includegraphics[width=0.99\hsize]{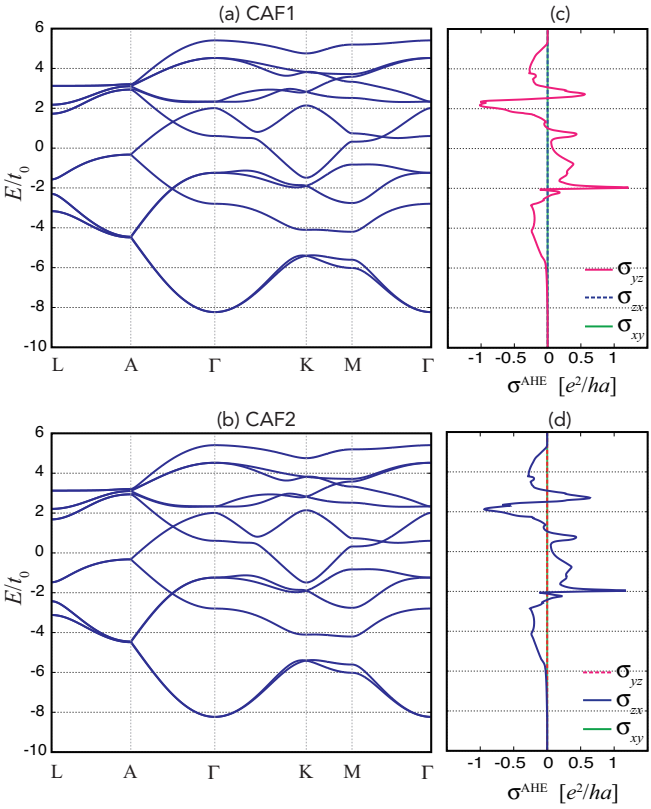} 
 \caption{
Band structure of the tight-binding Hamiltonian ${\cal H}(\vk)$ for configurations of local spins on Mn atoms (a) CAF1 and (b) CAF2, respectively. (c), (d)  The anomalous Hall conductivities as a function of the Fermi energy $E$ for (c) CAF1 and (d) CAF2.
 }
 \label{fig:Ek}
\end{figure}

The interaction between itinerant electron spins and localized spins is described by the exchange coupling term
\begin{equation}
  H_{\rm exc} =-J\sum_i \vm_i\cdot c^{\dag}_{i\sigma}\vs_{\sigma\sigma'} c_{i\sigma'}^{},
\end{equation}
where $\vm_i$ is the direction unit vector of the localized magnetic moment at $i$th site.
We focus on two cases with spin textures under the magnetic fields along the $x$ and $y$ directions, referred to CAF1 and CAF2, respectively, as shown in Fig. \ref{fig:lattice02} (a) and (b).
In terms of the Fourier components of the creation and annihilation operators, $c_{\vk\sigma}^{}=\frac{1}{\sqrt{N}}\sum_j c_{j\sigma}e^{i\vk\cdot{\bf R}_j}$, the total Hamiltonian is represented in the form of
\begin{eqnarray}
 H&=& H_0+H_{\rm so}+H_{\rm exc} \nonumber\\
&=&\sum_{\vk} c^{\dag}_{\vk\sigma}{\cal H}_{\sigma\sigma'}(\vk)c^{}_{\vk\sigma'}.
\label{H_total}
\end{eqnarray}
The energy eigenvalues, obtained by diagonalizing ${\cal H}(\vk)$, are shown along high-symmetry lines in Fig. \ref{fig:Ek} for (a) CAF1 and (b) CAF2 configurations.
Here we set $t_{\rm so}=0.2t_0$ and $J=1.7t_0$. These parameters have been estimated from the first principle calculation in Ref. \ref{Chen2014Ref} for Mn$_3$Ir. We checked that qualitative behaviors shown below remain unchanged as the parameters change.


{\it Anomalous Hall conductivity}---
Using the Kubo formula the anomalous Hall conductivity $\sigma_{ij}$ ($i\ne j$)  from the intrinsic mechanism is given by 
\begin{equation}
\sigma_{ij}=\frac{e^2}{h}\epsilon_{ijk}\sum_n\int_{BZ}\frac{\mathrm{d}^3\vk}{(2\pi)^2}\Omega_{n,k}(\vk)
f(E_{n\vk}-\mu),
\label{eq:AHC}
\end{equation}
where 
$\bm{\Omega}_n(\vk)=\bm{\nabla}_{\vk}\times\bm{A}_n(\vk)$
is the Berry curvature and 
$\bm{A}_n(\vk)=-i\left<u_{n\vk}\vline\bm{\nabla}_{\vk}\vline u_{n\vk}\right>$
is the Berry connection, $\vline\, u_{n\vk}\rangle$ being the Bloch state obtained by solving  the eigen equation 
$
{\cal H}(\vk)\,\vline\,
u_{n\vk}
\rangle=
E_{n\vk} \,\vline\, u_{n\vk} \rangle
$.
The anomalous Hall conductivities are shown as a function of the Fermi energy $E$ in Fig. \ref{fig:Ek} for (c) CAF1 and for (d) CAF2. In the CAF1 configuration $\sigma_{yz}$ is finite, while $\sigma_{zx}=\sigma_{xy}=0$.
In the CAF2 configuration, on the other hand, $\sigma_{zx}$ is finite, while $\sigma_{xy}=\sigma_{yz}=0$.
These results are consistent with experiments\cite{Nakatsuji2015,Nayak2015,Kiyohara2016,Li2017}.

\begin{figure}[b]
 \leavevmode\includegraphics[width=0.99\hsize]{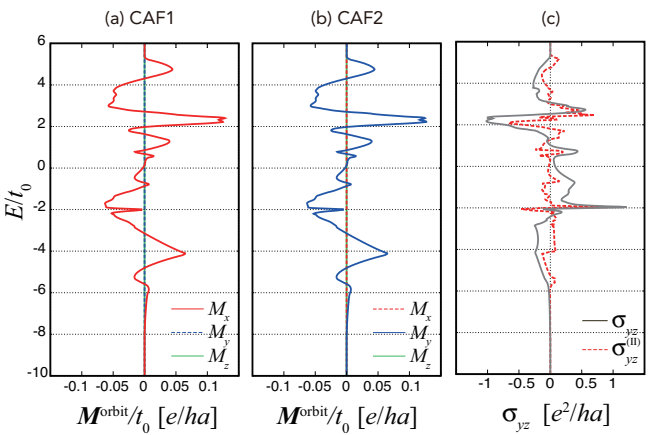} 
 \caption{
Orbital magnetization for (a) CAF1 and (b) CAF2 as a function of the Fermi energy $E$.
(c) Total anomalous Hall conductivity $\sigma_{yz}$ and $\sigma_{yz}^{(\rm I\!I)}$.
 }
 \label{energy}
\end{figure}


{\it Orbital magnetization}---
Next we consider the orbital magnetization which is obtained by the following equation\cite{Ceresoli2006,Xiao2010},
\begin{eqnarray}
 &&\vM^{\rm orbit}=\frac{e}{2h}\sum_n\int_{BZ}\frac{\mathrm{d}^3\vk}{(2\pi)^2}f(E_{n\vk}-\mu)
 \nonumber \\
 &&\qquad\quad \times i\left<\frac{\partial u_{n\vk}}{\partial\vk}\vline
 \times\left\{2\mu-E_{n\vk}-\mathcal{H}(\vk)\right\}\vline\frac{\partial u_{n\vk}}{\partial\vk}\right>,
\label{orbital magnetization}
\end{eqnarray}
and discuss the relation with the anomalous Hall effect.
Here $\mu$ is the Fermi energy.
%
The orbital magnetization $\vM^{\rm orbit}=(M_x,M_y,M_z)$ is shown in Fig. \ref{energy} (a) for CAF1 and (b) for CAF2 as a function of the Fermi energy $E$.
In CAF1 $M_x$ is finite, while $M_y=M_z=0$. In CAF2, on the other hand, $M_y$ is finite, while $M_x=M_z=0$.
These results show that the ground state of the our tight-binding Hamiltonian under the fixed local spin magnetizations (CAF1 and CAF2) possesses a finite orbital magnetization contrary to the fact that the net spin magnetization is zero.
In the presence of a magnetic field $\vB$ pointing in the Kagome plane, the orbital magnetization couples with the field\cite{Xiao2010,Ceresoli2006} via 
$-\vB\cdot\vM^{\rm orbit}$.
When $\vB$ points in the $x$ direction, the CAF1 configuration with $\vM^{\rm orbit}$ pointing in the $x$ direction is energetically favored.
Similarly, when $\vB$ points in the $y$ direction, the CAF2 configuration with $\vM^{\rm orbit}$ pointing in the $y$ direction is favored, consistent with experiments\cite{Nakatsuji2015,Nayak2015,Kiyohara2016,Li2017}.

\begin{figure}[t]
 \leavevmode\includegraphics[width=0.99\hsize]{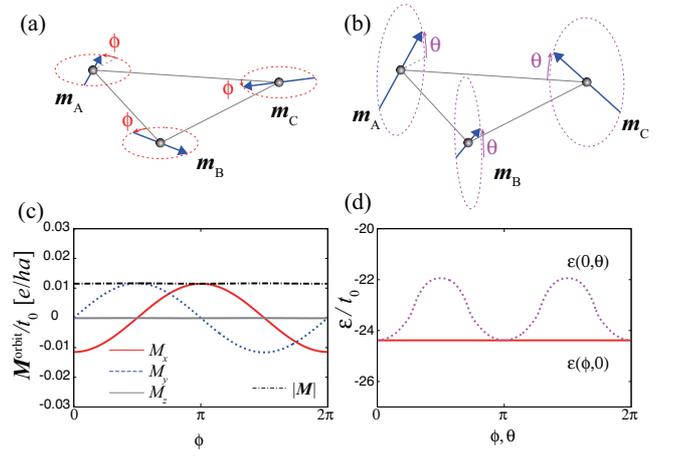} 
 \caption{
 Configurations of local spins on Mn atoms with (a) in-plane tilting angle $\phi$ and (b) out-of-plane tilting angle $\theta$.
 (c) Orbital magnetization as a function of in-plane tilting angle $\phi$. (d) Total energy of electrons as a function of $\phi$ and $\theta$.
 }
 \label{angle}
\end{figure}

Here we study the anisotropy of the orbital magnetization. We first consider the case where the local magnetizations are given as
$\vm_A=(\cos\phi,\sin\phi,0)$, $\vm_B=(\cos(\phi-\frac{2\pi}{3}),\sin(\phi-\frac{2\pi}{3}),0)$, 
$\vm_C=(\cos(\phi+\frac{2\pi}{3}),\sin(\phi+\frac{2\pi}{3}),0)$ on the A, B, and C sublattice, as shown in Fig. \ref{angle} (a).
Here $\phi$ is the in-plane tilting angle of the local spin magnetization from the CAF1 configuration.
Figure \ref{angle} (c) shows the angle dependence of the orbital magnetization $\vM^{\rm orbit}=(M_x, M_y, M_z)$ and $|\vM^{\rm orbit}|=\sqrt{M_x^2+M_y^2+M_z^2}$ as a function of $\phi$ at the Fermi energy $\mu=-t_0$. 
The results show that the orbital magnetization changes as $\vM^{\rm orbit}=M_0(-\cos\phi,\sin\phi,0)$, where the norm $M_0=|\vM^{\rm orbit}|$ is independent of $\phi$.
The qualitative behaviors remain unchanged as the Fermi energy changes.
To study the anisotropy of the orbital magnetization we compute the total energy of electrons,
$ \varepsilon=\frac{1}{N}\sum_{n,\vk}E_{n\vk}f(E_{n\vk}-\mu)$
as a function of $\phi$,
where $N$ is the number of unit cells and $f$ is the Fermi-Dirac distribution function. 
Figure \ref{angle} (d) shows the total energy of electrons as a function of $\phi$, $\varepsilon(\phi,0)$, at $\mu=-t_0$.
Now we compare $\varepsilon(\phi,0)$ to the energy in the case of out-of-plane local spin configuration
$\vm_A=(\cos\theta,0,\sin\theta)$, $\vm_B=(-\frac{1}{2}\cos\theta,-\frac{\sqrt{3}}{2}\cos\theta,\sin\theta)$, and $\vm_C=(-\frac{1}{2}\cos\theta,\frac{\sqrt{3}}{2}\cos\theta,\sin\theta)$, where $\theta$ is the out-of-plane tilting angle (Fig. \ref{angle} (b)).
As shown in Fig. \ref{angle} (d), the total energy computed as a function of $\theta$, $\varepsilon(0,\theta)$, has maxima at $\theta=\pi/2$ and $3\pi/2$, while $\varepsilon(\phi,0)$ is independent of $\phi$.
These behaviors clearly indicate that the system is an orbital ferromagnet with the easy-plane anisotropy.

By comparing the anomalous Hall conductivity and the orbital magnetization, $\sigma_{yz}$ and $M_x$ in CAF1 for example, one finds that both quantities are not proportional to each other. In the conventional ferromagnet, on the other hand, the anomalous Hall conductivity is proportional to the spin magnetization as seen in eq. (\ref{sigma_xy_FM}).

To see the connection between the anomalous Hall effect and the orbital magnetization,
we decompose the Hall conductivity into two terms\cite{Streda1982}
\begin{equation}
\sigma_{ij}=
 \sigma_{ij}^{\rm (I)} + \sigma_{ij}^{\rm (I\!I)}.
\label{eq:I+II}
\end{equation}
$\sigma_{ij}^{\rm (I)}=(e^2/h){\rm Tr}(v_iG^+v_jG^-)$, 
where $G^{\pm}=(\mu \pm i0-{\cal H})^{-1}$ and $v_i=\partial{\cal H}(\vk)/\partial(\hbar k_i)$, 
is associated with states on the Fermi surface.\cite{Streda1982} On the other hand, 
\begin{equation}
\sigma_{ij}^{\rm (I\!I)}
=-e \epsilon_{ijk} \frac{\partial M^{\rm orbit}_k}{\partial \mu}
\label{eq:streda}
\end{equation}
is the contribution of all filled states below the Fermi energy and is a thermodynamic equilibrium property of the magnetic material.\cite{Streda1982}
In fig. \ref{energy} (c) the anomalous Hall conductivity $\sigma_{yz}$ obtained by eq. (\ref{eq:AHC}) and $\sigma_{yz}^{\rm (I\!I)}$, eq.(\ref{eq:streda}), are compared. Near the peaks of the anomalous Hall conductivity, these quantities are close each other.

\begin{figure}[t]
 \leavevmode\includegraphics[width=0.99\hsize]{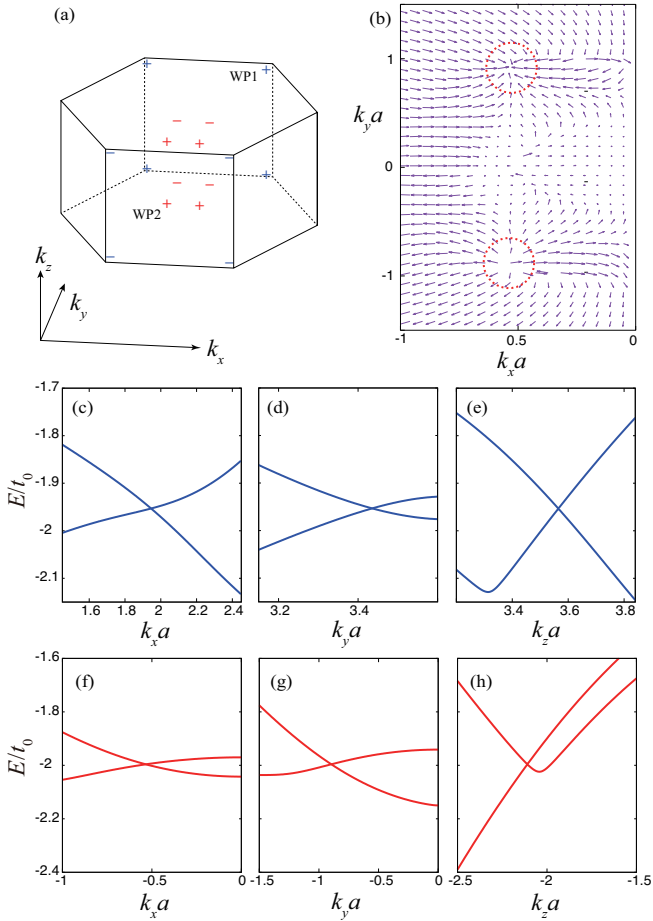} 
 \caption{
 (a) Positions of Weyl points in momentum space. 
 The sign $\pm$ corresponds to the chiralities of the Weyl points.
 (b) Direction of the Berry curvature around the Weyl points. Dotted circles indicate the positions of the Weyl points.
(c)-(h) Energy dispersions around the Weyl points WP1 and WP2.  }
 \label{WP}
\end{figure}


{\it Weyl points}---
Finally we study the Weyl points in the spectrum of ${\cal H}(\vk)$.
Weyl semimetals and Weyl metals are characterized by the presence of nondegenerate band-touching points (Weyl points) which appear as separated pairs in momentum space\cite{Wan2011,Burkov2011}.
In terms of the Berry curvature, Weyl points are synthetic magnetic monopoles, and thus contribute to the anomalous Hall conductivity.
Here we focus on the Weyl points appearing between fourth band and fifth band in CAF2, where the anomalous Hall conductivity $\sigma_{zx}$ takes a maximum value around $E/t_0=-2$ in Fig. \ref{fig:Ek} (d).
The model Hamiltonian ${\cal H}(\vk)$ with parameters shown above has 16 Weyl points between fourth and fifth bands in the first Brillouin zone as shown in Fig. \ref{WP} (a). 
Figure  \ref{WP} (b) shows the Berry curvature\cite{Xiao2010} of the fourth band
 in the vicinity of the Weyl point WP2 noted in  Fig. \ref{WP} (a).
 By examining whether the Berry flux is inward or outward, the chirality of the Weyl points can be assigned by the sign $\pm$ in Fig. \ref{WP} (a).
Figures \ref{WP} (c)-(h) show energy spectra near the Weyl points.

We now present a qualitative argument showing the relation between the directions of the orbital magnetization, anomalous Hall conductivity tensor, and the positions of the Weyl points.
In a Weyl semimetal
the low-energy excitations are described by the Weyl Hamiltonian\cite{Wan2011,Burkov2011}
\begin{equation}
{\cal H}_{\rm Weyl}^{(\alpha)\pm}(\vk)
=\pm \hbar v_F\vs \cdot(\vk - \vQ^{(\alpha)}_\pm).
\label{eq:weyl}
\end{equation}
Here we consider the case where the energy dispersion is isotropic for simplicity.
The Weyl points appear as pairs with opposite chiralities: the sign in the right hand side of eq. (\ref{eq:weyl}).
$\vQ^{(\alpha)}_+$ is the position of the Weyl point of $\alpha$th pair with the positive chirality, and $\vQ^{(\alpha)}_-$ is that with the negative chirality.
The separation of the Weyl points $\Delta\vQ^{(\alpha)}=\vQ^{(\alpha)}_+-\vQ^{(\alpha)}_-$ can be realized when time-reversal and/or inversion symmetries are broken.
When time-reversal symmetry is broken, the anomalous Hall effect occurs. 
In an ideal Weyl semimetal where all Weyl points appear at the same energy, say $\mu=0$, the anomalous Hall conductivity is given by the sum of contributions from pairs of Weyl points separated by $\Delta{\bm Q}^{(\alpha)}$,
\begin{equation}
  \sigma^{\rm Weyl}_{ij}=\epsilon_{ijk}\frac{e^2}{2\pi h}\sum_{\alpha}\Delta Q_k^{(\alpha)},
\label{AHC_Weyl}
\end{equation}
at $\mu=0$.\cite{Wan2011,Burkov2011}
In terms of decomposition of the anomalous Hall conductivity eq. (\ref{eq:I+II}), $\sigma^{({\rm I})}_{ij}=0$ because there is no finite Fermi surface at $\mu=0$, vanishing the density of states.
On the other hand, eq. (\ref{eq:streda}) connects the Hall conductivity and the orbital magnetization, leading the relation \cite{note}
\begin{equation}
  \vM^{\rm orbit} = -\frac{e \mu}{2\pi h} \sum_{\alpha} \Delta\vQ^{(\alpha)}
\label{M_Weyl}
\end{equation}
at the Fermi level $\mu$.

In our model the band dispersions are bent as shown in Fig. \ref{WP} so that hole pockets appear when the Fermi level resides at the Weyl point. The anomalous Hall conductivity and the orbital magnetizations are not as simple as eqs. (\ref{AHC_Weyl}) and (\ref{M_Weyl}). Nevertheless, qualitative behaviors such as the relation between the directions of the orbital magnetization, the anomalous Hall conductivity tensors, and the configurations of the Weyl points are the same as those.
Indeed, the position of the Weyl points shown in Fig. \ref{WP} (a) are $a^{-1}(\pm0.5341,\pm0.8948,\pm2.110)$ and $a^{-1}(\pm1.948,\pm3.434,\pm3.565)$, and the sum of the relative displacements of all pairs, $\sum_{\alpha}\Delta\vQ^{(\alpha)}=a^{-1}(0,8.71,0)$, point the $y$ direction. Thus, eq. (\ref{AHC_Weyl}) indicates that $\sigma_{zx}\ne0$ and $\sigma_{xy}=\sigma_{yz}=0$, and eq. (\ref{M_Weyl}) indicates that $M_y\ne0$ and $M_x=M_z=0$, consistent to the results shown in Figs. \ref{fig:Ek} and \ref{energy}. 
The estimated value $\sigma^{\rm Weyl}_{zx}=1.39\ [e^2/ha]$ and $\partial M_y/\partial \mu=-1.39\ [e/ha]$, from eq. (\ref{AHC_Weyl}) and eq.(\ref{M_Weyl}) respectively,
are in reasonable agreement with the anomalous Hall conductivity $\sigma_{zx}=1.21\ [e^2/ha]$
and $\partial M_y/\partial \mu=-1.16\ [e/ha]$
at $E=-1.96t_0$ in the lattice model shown in Figs. \ref{fig:Ek} and \ref{energy}.



{\it Conclusion}---
In this work we studied the anomalous Hall effect and the spontaneous orbital magnetization in chiral antiferromagnets.
We showed that the single-band tight-binding Hamiltonian indicates an orbital ferromagnetic order with the easy-plane anisotropy.
The symmetry relation between the anomalous Hall conductivity and the local spin configuration is 
characterized by the Weyl points and is consistent with experiments. 

While we were finalizing the paper, a preprint Ref. \ref{Li2017Ref} appeared. While we focus on itinerant electrons on Mn atoms, Ref. \ref{Li2017Ref} focuses on $Z$ atoms in Mn$_3Z$ where $Z$=Sn or Ge.


 This work was supported by JSPS KAKENHI Grants No. JP15H05854 and No. JP26400308.

\end{document}